# Impacts of Typhoons on Kuroshio Large Meander: Observation Evidences

Liang SUN[1,2], Yuan-Jian Yang[1], and Yun-Fei Fu[1]

1. Laboratory of Satellite Remote Sensing and Climate Environment, School of Earth and Space Sciences, University of Science and Technology of China, Hefei, Anhui, 230026, China;

2. LASG, Institute of Atmospheric Physics, Chinese Academy of Sciences, Beijing 100029, China

Corresponding author address:

Liang SUN

School of Earth and Space Sciences, University of Science and Technology of China

Hefei, Anhui 230026, China

Phone: 86-551-3600175; Fax: 86-551-3606459

Email: sunl@ustc.edu.cn; sunl@ustc.edu

**Abstract:**

The formation of the Kuroshio large meander in summer 2004 was investigated by using the cruise data, Argo profiles data, and satellite remote sensing data. We validated the point that cyclonic eddy contributes to the large meander. Besides, the impacts of typhoons on Kuroshio meanders were studied. From 29 July to 4 August, the typhoons stirred the ocean and upwelled the deep water, which enhanced the existed cyclonic eddy, and immediately made a drastic meander of the Kuroshio. Moreover, the unexpected typhoons in June 2004 also contributed to the initial meander at Tokara Strait.

The result suggests an alternative meander mechanism of Kuroshio path via typhoon-eddy-Kuroshio interactions. It is argued that typhoons accompanied with cyclonic eddies, might play crucial roles in meanders of the Kuroshio. This will provide a more comprehensive understanding of the dynamics of the west boundary flows like the Kuroshio and the Gulf Stream, and will be useful in eddy-resolution models.

## 1. Introduction

When the Kuroshio flows in Shikoku Basin, the sea south of Japan, it might have three different paths: the nearshore non-large-meander path, the offshore non-large-meander path and the large-meander path [Kawabe, 1985, 1995; Akitomo et al., 1991; Miyazawa et al., 2004, 2008]. The shift from one path to another would significantly cause oceanic changes and correspondingly climatic changes due to the strong eddy-current interactions and air-sea interactions at the sea surface along the Kuroshio path [Qiu, 2002, 2003; Taguchi et al. 2005]. Besides, the variations of both shape and position of the current also have large influences on fisheries, ship navigation, marine resource, etc [Kawabe, 1985; Taguchi et al., 2005].

Though the Kuroshio is basically driven by wind stress curl over the North Pacific, the mechanisms of Kuroshio path meander are mainly attributed to ocean dynamics, including local topography [Taguchi et al., 2005], upstream volume transport at Tokara Strait [Akitomo et al., 1991, 1996; Kawabe, 1995; Nakamura et al., 2006] and low potential-vorticity (PV) water at Shikoku Basin [Qiu and Miao 2000; Taguchi et al., 2005]. To understand the large meander of the Kuroshio path, the processes are divided into two parts: the trigger process and the formation process [Kawabe, 1995], which have been simulated by ocean models. First, there is a trigger meander at Tokara Strait, which can be either a large inflow [Akitomo et al., 1991; Nakamura et al., 2006], a small meander of path [Kawabe, 1995; Miyazawa et al., 2008], low PV water [Qiu and Miao 2000], or high PV water [Akitomo and Kurogi, 2001]. Then the initial meander is advected downstream to Shikoku Basin, where the large meander of Kuroshio path eventually occurs.

Recently, the Kuroshio took the large meander path from summer 2004 to summer 2005 after a non-large meander period of 13 years [Usui et al., 2008]. The trigger and formation of large meander formation were analyzed in JCOPE (the Japan Coastal Ocean Predictability Experiment) ocean forecast system [Miyazawa et al., 2008], which is a high-resolution ocean model [Kagimoto et al., 2008] with assimilated observation data including the sea surface height anomaly (SSHA), sea surface temperature (SST), and temperature/salinity profiles, etc. Due to the high performance of the forecast system, it can simulate the real path meander south of Japan in some sense.

However, differences were noted between the simulated Kuroshio path and the observations. For example, the model skill is worse than both the persistence and the climatology during the period from 21 May to 21 June [Miyazawa et al., 2008]. And the simulated meander in July 2004 was notably smaller than the observations. Moreover, both numerical simulations have not yet investigated the larger meander formation in August 2004, when the Kuroshio path had a drastic larger meander, even larger than that in July 2004.

To understand what happened in August 2004, the trigger force beyond oceanic processes is considered. It is noted that there were many typhoons in 2004, which passed the East China Sea and Tokara Strait. At the sea Tokara Strait, two typhoons blew across the water in June 2004 (Fig. 1a). Besides, from July 29th to August 4th, 2004, there were two consecutive typhoons (Namtheun and Malou) passing over the sea south of Japan. Meanwhile, there was a cyclonic eddy right near the typhoon track. The vigorous air-sea interactions induced vertical mixing and strong upwelling, which eventually changed Kuroshio path. In this research, the variations of the Kuroshio axis are examined before and after typhoons Namtheun and Malou. Therefore, the impacts of typhoons on the meander of the Kuroshio are discussed.

**2. Data and methods**

Sea surface wind vector and stress with spatial resolution of 0.25°×0.25°, were obtained from the daily QuikSCAT (Quick Scatterometer) provided by the Remote Sensing Systems (http://www.remss.com/). The wind stress was calculated with the bulk formula [Garrett, 1977], and the upwelling due to wind was calculated by using the Ekman pumping formula [Price et al., 1994]. Altimeter data were derived from multi-sensors. These sensors were Jason-1, TOPEX/POSEIDON, GFO, ERS-2 and Envisat. Data were produced and distributed by AVSIO (Archiving, Validation and Interpretation of Satellite Oceanographic data). Near-real time merged (TOPEX/POSEIDON or Jason-1 + ERS-1/2 or Envisat) sea surface height anomaly (SSHA) data, which are high resolution of 1/3°×1/3° Mercator grid, are available at www.aviso.oceanobs.com. The Kuroshio axis data are from the weekly Quick Bulletin Ocean Conditions provided by the Hydrographic and Oceanographic Department of the Japan Coastal Guard (JCG). Vertical profile data for temperature at cruise stations on 07/16 and 08/10 were from Japan Oceanographic Data Center (JODC). And the vertical profiles of float 2900320 from 07/16 to 08/13 were extracted from the real-time quality controlled Argo data base of China Argo Real-time Data center.

## 3. Results

*3.1 Role of cyclonic eddy in Kuroshio path*

Figure 1b shows the SSHA and the Kuroshio axis form June 23, 2004 to July 20, 2004. There was a large cyclonic eddy in the sea south of Japan, with recirculation gyre center located about at 32°N. The Kuroshio axis, therefore, had a significant meander around the recirculation gyre, where the commonly east-north-ward slant axis turned to south-north-ward straight axis at 135.5°E. As there was no significant anticyclonic eddy, it seemed that the northern cyclonic eddy occupied the nearshore region, and extruded the Kuroshio axis offshore, which made a right protrusion of the Kuroshio axis. During the time, the cyclonic eddy was advected eastward downstream with the Kuroshio, and there was an anticyclonic eddy emerging at south of the Kuroshio path. The anticyclonic eddy then pushed the Kuroshio close to the sea east of Kyushu, which made the Kuroshio axis straightly along 32°N. Meanwhile, the straight Kuroshio axis moved eastward with the advection of cyclonic eddy along 32°N from 135.5°E to 136.9°E, about 80 km away from its original position.

The above results prove that the large cyclonic eddy plays an important role in the Kuroshio path. The similar results were previously obtained by model simulations [Akitomo and Kurogi, 2001], and then by the observations [Ebuchi and Hanawa 2003]. The recent model simulation also noted that the cyclonic eddy contributes to large meander [Miyazawa et al., 2008]. Here, the observations confirmed such conclusion that large cyclonic eddy plays an important role in the meander of the Kuroshio path.

*3.2 Impact of typhoons on large meanders*

From 29 July 2004, two typhoons consecutively passed over the sea south of Japan (Fig. 1a). They stirred the surface sea water and upwelled deep sea water. Correspondingly, there was strong upwelling near the region at 32°N, 137°E due to the strong winds, where the maximum upwelling velocity was over $12 \times 10^{-4}$ m/s, which is more than 3 times that in model simulations [Miyazawa et al., 2008]. Meanwhile, there was a large weakly downwelling region around the upwelling area. Compared the position of the cyclonic with that of typhoon track, it is clearly that the strong upwelling occurred at right around the cyclonic eddy region. This can be seen from Fig 2b, where the vertical profiles of temperature at cruise stations (in the cyclonic eddy) significantly cooled after the typhoon's passage. And the temperature of water beneath 100 m became little warmer than before at Argo float (in the anti-cyclonic eddy) due to downwelling of surface warm water. However, there was also vertical mixing due to upwelling and downwelling processes, which cooled the surface temperature and warmed the water beneath 50 m (solid curves in Fig 2b). Furthermore, the strong upwelling also made the strong sea surface cooling, and accordingly the phytoplankton bloom (not shown).

In consequence, the cyclonic eddy enhanced due to the upwelling by typhoons, and played the major role in the formation of the large meander (Fig. 2c), where the Kuroshio axis took a larger meander shortly after typhoons passage. Compared to the previous state on July 23 2004, the cyclonic eddy enlarged and intensified due to strong upwelling on 6 August 2004 shortly after the typhoons passage. It pushed the Kuroshio path to the south, and immediately made a right shift of the Kuroshio path. The Kuroshio path bent consequently around the cyclonic eddy. The bend continued till the Kuroshio path was around the enhanced recirculation gyre totally on 13 August 2004, which made the largest meander of the Kuroshio path with about more that 100 km on 27 August 2004. The enhanced recirculation gyre was so strong that this large meander persisted till October 2004, when another extremely large typhoon Ma-on passed the same region again (not shown). The large cyclonic eddy then enhanced again, which made the large meander last for about one year until the cyclonic eddy declined due to dissipation.

To summarize, the typhoons, when they pass over cyclonic eddies, have strong air-sea interaction, and successively enhance the cyclonic eddies due to upwelling. The enhanced cyclonic eddies bend the Kuroshio path, which make large meander formation eventually. This implies that typhoons might have great impacts on the meander of the Kuroshio path via cyclonic eddies.

## 4. Discussion

As mentioned above, the model skill is worst during the period from 21 May to 21 June [Miyazawa et al., 2008]. Our investigation here points out that two typhoons in June 2004 that caused the Kuroshio axis meandering might be able to explain the poor performance of the model skill. According to the wind stresses by typhoons Conson and Dianmu, there was strong upwelling occurring at Tokara Strait and the sea south of Shikoku (Fig. 3a). The cyclonic eddy enhanced due to this, which consequently changed the Kuroshio axis (Fig. 3b). As the Kuroshio axis took changes due to typhoons, this might also be one of the trigger effects of large meander formation unresolved by the oceanic model. The unexpected typhoons made the model skill become worse than before.

Although typhoons might trigger initial small meanders, we should point out that there have been initial meander before typhoons' passages, and that the amplitude of initial meander was quite large, which can be seen from Fig. 3b. The investigations of long-range initial small meander were focused on the anticyclonic eddies [Akitomo, 1996; Qiu and Miao, 2000; Ichikawa, 2001; Endoh and Hibiya, 2001]. For the large meander in 2004, the investigations traced the initial small meander back from October 2003 to April 2004 [Usui et al., 2008; Miyazawa et al., 2008]. It is concluded that the strong anticyclonic eddies lead to the amplification of the trigger meander, and that the initial meander must be large enough to trigger the large meander [Usui et al., 2008; Miyazawa et al., 2008]. It seems that the initial small meander, accompanied with typhoon induced upwelling, triggered the large meander on July 2004.

Moreover, not all the typhoons played crucial roles in the large meander. For example, the typhoons Conson and Malou played minor roles in this case, which can be seen from both the upwelling and the Kuroshio axis (Fig. 2 and Fig. 3). Only when the typhoons have great interactions with eddies, the impacts of typhoons might make sense. This makes the forecast more difficult, as the predictions of typhoon tracks are not well solved yet.

The above result clearly showed the impact of typhoons on formation of the Kuroshio meander. There are other factors which are also necessary for the formation of the Kuroshio meander. First is the cyclonic and/or anticyclonic eddy in the sea south of Japan [Qiu and Miao 2000; Akitomo and Kurogi, 2001; Endoh and Hibiya, 2001; Ebuchi and Hanawa, 2003; Waseda et al., 2005]. Different from some previous studies [Waseda et al., 2005], the cyclonic eddy more than anti-cyclonic eddy played key role in our study according to the observations, which agrees with other numerical simulations [Miyazawa et al., 2008]. Secondly, the Izu Ridge, which produces a cyclonic torque over the western slope of the ridge when the flow impinges upon it, is important for blocking the Kuroshio large meander from propagating eastward across the ridge [Mitsudera et al., 2006]. Third, the Kuroshio volume transport at Tokara Strait was notably larger on April and May 2004 from the observation [Andres et al., 2008], which might be the original perturbation on April 2004 [Miyazawa et al., 2008].

In the North Atlantic, there are Gulf Stream system and hurricanes [Sriver and Huber, 2007], which are similar to the Kuroshio system and typhoons in the North Pacific. The variations of the Gulf Stream were also recognized mainly as oceanic processes [Dijkstra, 2005]. Although there were some works discussed the effects of the strong atmospheric disturbances on the Gulf Stream or Kuroshio [Xue and Bane, 1997; Miyazawa and Minato, 2000, Wu et al., 2008]. Our results here also imply that hurricanes might also be important for the variations of the Gulf Stream, like that typhoons for the Kuroshio meanders.

In a word, typhoons can trigger the formation of the Kuroshio meander via typhoon-eddy-Kuroshio interactions, which is an alternative mechanism of Kursoshio meander. It is argued that not only oceanic processes, but also typhoons can have significant influence on the Kuroshio system. This implies that to accurately predict the Kuroshio path, the large synoptic processes, especially tropic cyclones, should also be considered.

## 5. Conclusion

In summary, the impacts of typhoons on the formation of the Kuroshio large meander in summer 2004 are investigated via observational data. We firstly confirmed the former conclusion that cyclonic eddy contributes to the large meander. Moreover, using the observation data, it was found that the cyclonic eddy accompanied with typhoons is the major factor of the Kuroshio large meander formation in summer 2004. From 29 July to 4 August, the typhoons stirred the ocean and upwelled the deep sea water, which enhanced the existed cyclonic eddy. The enhanced cyclonic eddy pushed the Kuroshio path to the south, and immediately made a right shift of the Kuroshio path with more that 100 km. This large meander of the Kuroshio path existed for about 1 year.

Then the impact of typhoons on the initial small meander was also discussed. It was found that the unexpected typhoons in June 2004 affected the model skill to be worse than before, and that they also contributed to the initial meander at Tokara Strait. In this case, the pre-existed eddy was indispensable, as the most vigorous air-sea interaction due to typhoons occurred always near the pre-existed eddies.

The results suggest an alternative meander mechanism of Kuroshio path via typhoon-eddy-Kuroshio interactions. As only the oceanic processes (inflow velocity, mesoscale eddies and local topography) were proposed to be the triggers of Kursohio meander, it is argued that typhoons accompanied with cyclonic eddies, might also be crucial in the Kuroshio path meander. It implies that to accurately predict the Kuroshio path, the large synoptic processes, especially the typhoons, should be taken into account at least. This will be likely to provide a more comprehensive understanding of the dynamics of the west boundary flows like the Kuroshio and the Gulf Stream, and will be useful in the numerical models, especially in eddy-resolution models.

**Acknowledgements:**

This work is supported by the National Basic Research Program of China (No. 2007CB816004), the National Foundation of Natural Science (Nos. 40705027, 40730950 and 40675027), and the Program for New Century Excellent Talents in University. We thank STI for providing typhoon track data, Remote Sensing Systems for QuikScat wind-vector data, AVISO for SSHA data, JCG for the Kuroshio axis data, JODC for cruise station data, and China Argo Real-time Data center for float profiles.

Figure 1. (a) The study region and the typhoons' tracks, 08/01 representing August $1^{st}$, etc. (b) The SSHA and the Kuroshio axis (bold curve). The cyclonic eddy played crucial role in the Kuroshio path.

Figure 2. (a) The wind stress (arrows) and upwelling (filled) during the typhoons' passages from 07/29 to 08/04, and the notation 06/30-07/02 presents the average value from 06/30 to 07/02, etc. (b) The SSHA (shadowed) and the locations of the cruise stations and Argo float (left); the vertical profiles of temperature at cruise stations (on 07/16 and 08/10) and Argo float (07/16 to 08/13) pre and post typhoon (right). (c) The SSHA and the Kuroshio axis (bold curve) before and after typhoons Namtheun and Malou from 07/21 to 08/27, and the notation 07/21-23 represents the average value from 07/21 to 07/23, etc.

Figure 3. (a) The wind stress (arrows) and upwelling (filled) during the typhoons' passages from 05/11-05/21. (b)

1. The SSHA and the Kuroshio axis (bold curve) before and after typhoons Namtheun and Malou from 05/26 to 06/28, and the notation 05/26-28 represents the average value from 05/26 to 05/28, etc

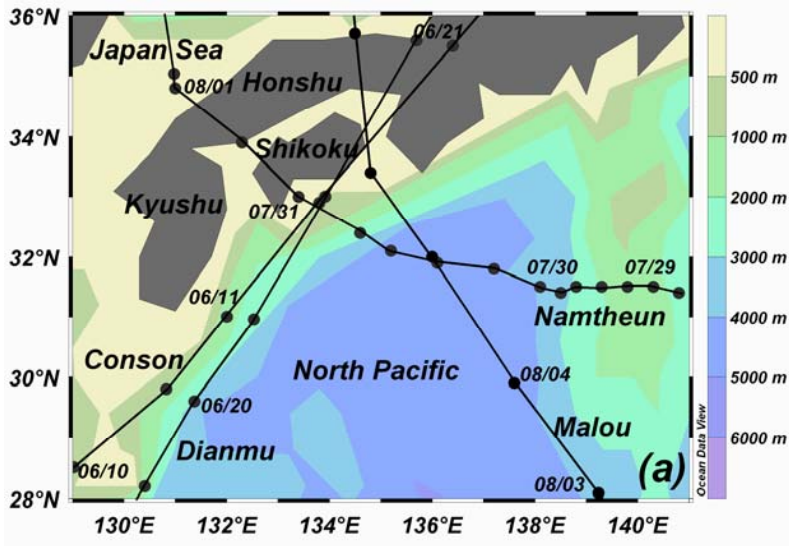

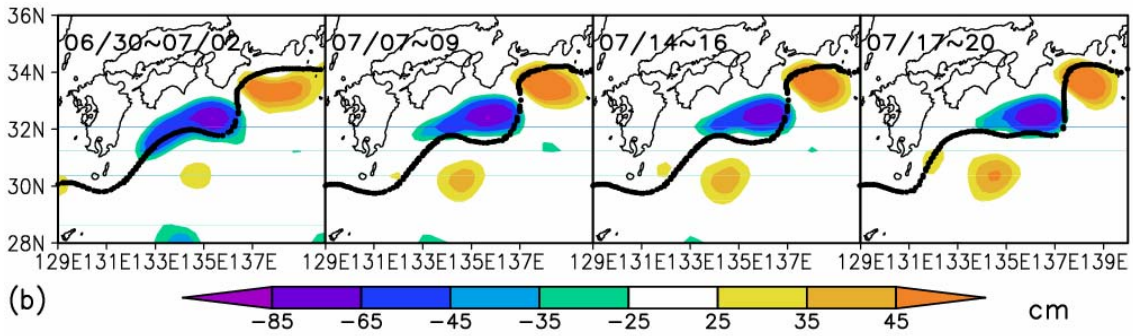

Figure 1

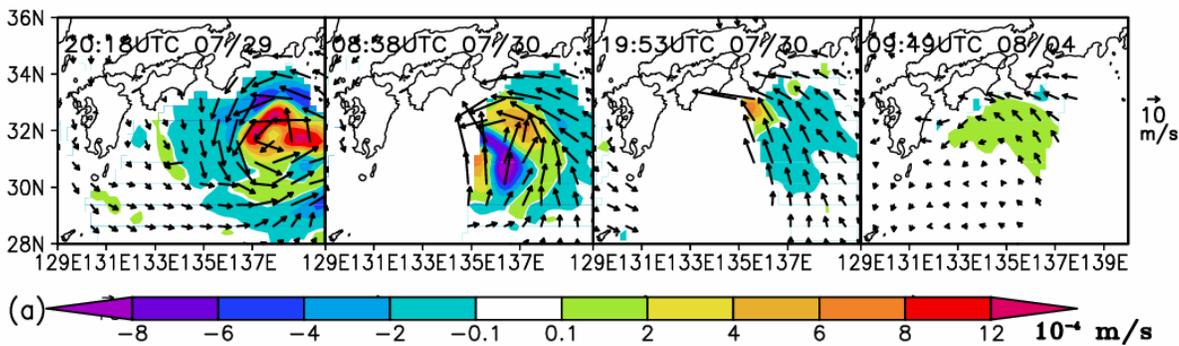

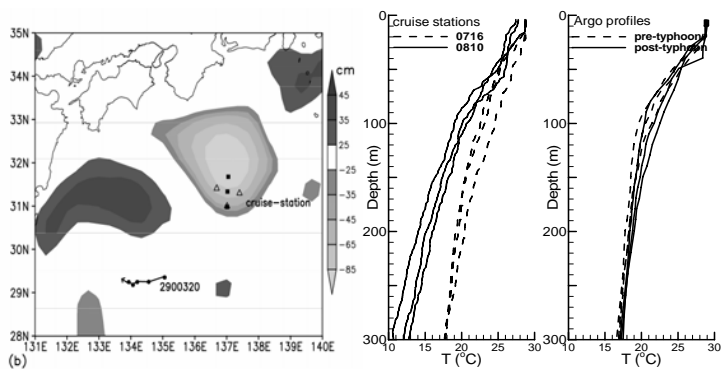

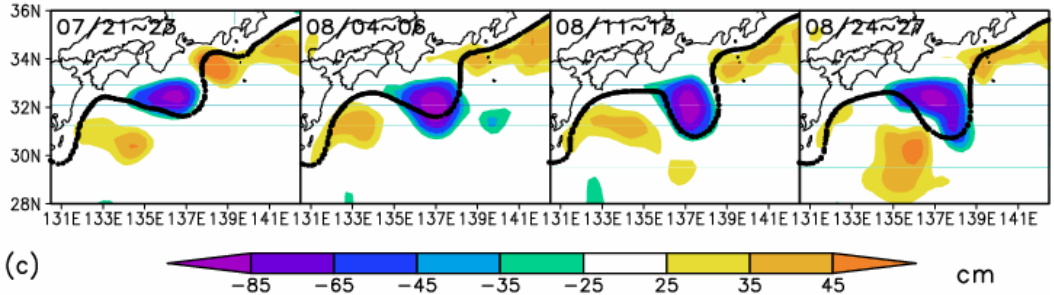

Figure 2.

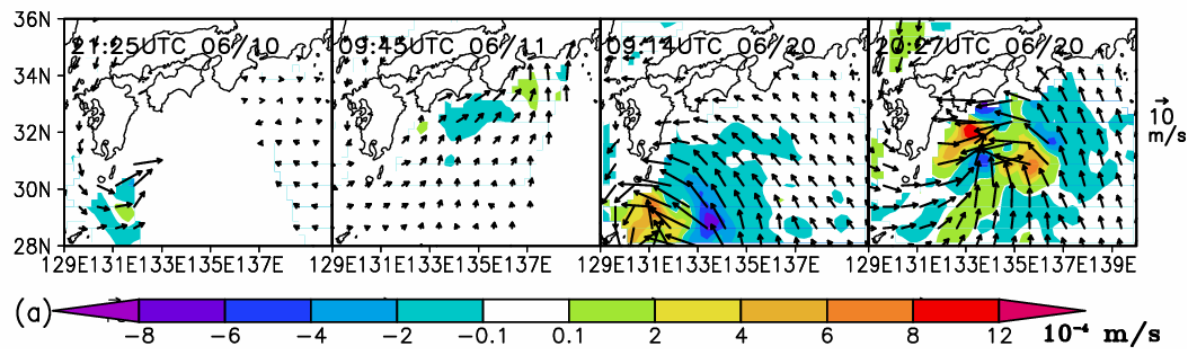

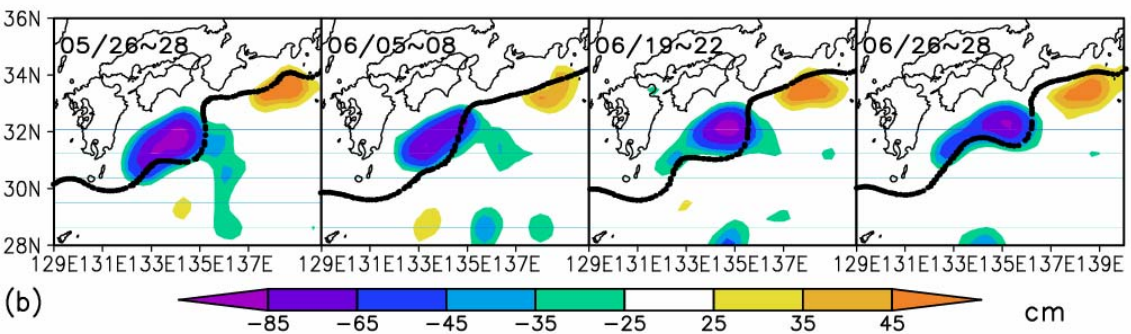

Figure 3.

14